\documentclass{emulateapj}

\usepackage{color}
\usepackage[normalem]{ulem}

\newcommand{\pd}[2]{\frac{\partial #1}{\partial #2}}
\newcommand{\pdd}[2]{\frac{\partial^2 #1}{\partial #2^2}}
\newcommand{\Mm}{~\mathrm{Mm}}
\newcommand{\km}{~\mathrm{km}}
\begin{document}

\title{Spectropolarimetrically accurate magnetohydrostatic sunspot model for forward modelling in helioseismology}
\author{D. Przybylski, S. Shelyag, P.S. Cally}
\shortauthors{Przybylski \it{et al.}}
\shorttitle{Radiatively accurate magnetohydrostatic sunspot model}
\affil{Monash Centre for Astrophysics, School of Mathematical Sciences, Monash University, Clayton, Victoria, Australia, 3800.}

\begin{abstract}
We present a technique to construct a spectropolarimetrically accurate magneto-hydrostatic model of a large-scale solar magnetic field concentration, mimicking a sunspot.
Using the constructed model we perform a simulation of acoustic wave propagation, conversion and absorption in the solar interior and photosphere with the sunspot embedded into it.
With the $6173\mathrm{\AA}$ magnetically sensitive photospheric absorption line of neutral iron, we calculate observable quantities such as continuum intensities, Doppler velocities, 
as well as full Stokes vector for the simulation at various positions at the solar disk, and analyse the influence of non-locality of radiative transport in the solar photosphere on helioseismic measurements. Bisector shapes were used to perform multi-height observations. The differences in acoustic power at different heights within the line formation region at different positions at the solar disk were simulated and characterised. An increase in acoustic power in the simulated observations of the sunspot umbra away from the solar disk centre was confirmed as the slow magneto-acoustic wave. 

\end{abstract}

\section{Introduction}

    Techniques of local helioseismology are currently unable to unambiguously determine sub-surface structure of the flows and sound speed perturbations in and around large-scale solar magnetic field concentrations, such as sunspots and pores \citep{shel_2007_AAa,gizo_2009_SSRv,mora_2010_SoPh}. Due to the complexity of magnetohydrodynamic processes involved, our understanding of the behaviour of magnetoacoustic waves as they are absorbed, reflected and refracted by sunspots is far from complete. 
    
Recently, it was demonstrated that solar magnetic fields and the process of magneto-acoustic wave mode conversion associated with them lead to significant changes in the wave travel times used in helioseismic inversions \citep{mora_2014_ApJ,hans_2014_SoPh}. Four processes associated with strong magnetism dominate wave behaviour in sunspots \citep{call_2015_AGU}: fast/slow mode conversion at the Alfv\'en/acoustic equipartition level $v_a=c_s$, allowing acoustic (slow) waves to transmit into the upper atmosphere if the `attack angle' $\alpha$ between wave vector and magnetic field is small but converting them to magnetic (fast) waves otherwise \citep{call_2006_RSPTA,schu_2006_MNRAS}; the ``ramp effect'' that reduces the effective acoustic cutoff frequency $\omega_c$ to $\omega_c \cos\theta$, where $\theta$ is the magnetic field inclination from the vertical \citep{bel_1977_AA}; fast wave reflection around the height where the Alfv\'en speed matches the wave's horizontal phase speed; and fast/Alfv\'en mode conversion that typically occurs over several scale heights near the fast wave reflection level, generating both upward and downward propagating Alfv\'en waves \citep{call_2011_ApJ}. Fast/slow conversion is found to produce large negative travel time shifts, while fast/Alfv\'en conversion generates countervailing positive shifts provided the vertical plane containing the wave vector is nearly perpendicular to the vertical plane containing the magnetic field lines \citep{call_2013_MNRAS}. The ramp effect allows field-guided acoustic waves to enter the atmosphere in inclined field where frequency $\omega>\omega_c \cos\theta$ while normally the acoustic cutoff would prevent their propagation ($\omega<\omega_c$). The complexity and sensitivity to magnetic field direction of wave motions above the equipartition level makes interpretation of observations difficult but potentially rewarding. Moradi et. al. (2015, submitted) studied the effects of directional time-distance helioseismology on the travel time measurements in the sunspot model. 

There are also possible significant discrepancies in travel time measurements originating from the effects of non-locality of radiative transport in the solar atmosphere. Changes in spectral line formation heights due to magnetic field presence \citep[see e.g.][]{shel_2007_AA}, systematic centre-to-limb variations in absorption line formation \citep{shel_2014_PASJ}, as well as instrumental effects, such as stray light, and other processes involved in formation and measurement of radiation intensities and Doppler shifts result in our inability to unambiguously measure the travel time perturbations and, therefore, infer solar sub-surface structure \citep{raj_2011_ASInC}.

Rapid improvements in computational power already make it possible to perform forward modelling of magnetohydrodynamic wave propagation and mode conversion in ``realistic" solar magnetic field structures \citep{shel_2009_AA, mora_2009_ApJ,feli_2010_ApJ,cam_2011_SoPh,khom_2012_ApJ,feli_2012_ApJ,zhar_2013_AnGeo,feli_2014_ApJ}. Spectral line synthesis codes and radiative diagnostics tools also allow computations of mock observables from the simulated plasma parameters, allowing for direct comparison between simulations and observations in computational helioseismology. 

Creating a sunspot that is both spectropolarimetrically accurate and magnetohydrostatically stable is inherently difficult, as the sound speed and temperature can change significantly with small changes in the density and pressure stratification. The sunspot model of \citet{khom_2008_apj} was created to allow empirical quiet and umbral solar models to be used in the near-surface layers in combination with a Schl\"uter-Temesvary flux tube model \citep{scl_1958_IAUS} in the interior. However, the model created this way is still not convectively stable. Convective instability is fatal to linear MHD simulations, but these codes are less expensive than full non-linear simulations, and ideal for the long time series required in helioseismology; for the study of fast and slow magneto-acoustic waves; and for simulating fast-slow and fast-Alfv\'en mode conversion in the photosphere and lower chromosphere. The effects of convective stabilisation on the eigenmodes of solar models for helioseismic simulations were studied by \citet{schu_2011_SoPh}. A technique for stabilising the atmosphere is discussed in Sec.~\ref{sec:model}.

In this paper, we present a model of a magneto-hydrostatic and spectropolarimetrically accurate sunspot. Our model is based on the sunspot-like model of \citet{khom_2008_apj}. The model was adjusted to provide a more accurate replication of photospheric sunspot properties taken from semi-empirical models, while still maintaining a smooth transition of physical properties between the magnetic and non-magnetic regions required for stable numerical simulation. This technique makes it possible to obtain accurate photospheric absorption line formation heights as well as allowing the study of observational signatures of acoustic wave propagation in the simulated model at different positions on the solar disk. We perform a magnetohydrodynamic simulation of the propagation of a wave through this sunspot-like model and investigate the behaviour of acoustic waves in the simulated model using the synthesised radiation, as if it were observed. We also investigate effects of the centre-to-limb variation effects on Doppler velocity measurements and study the line bisector shapes to allow for a multi-height view in the line formation region, which can be used to observationally disentangle wave mode conversion process in the solar atmosphere. 

The structure of the paper is as follows. In Section 2 we describe the background model. In Section 3 we explain the magnetohydrodynamic simulation and the spectral synthesis methods used to provide artificial observables. Section 4 provides results and description of the radiative effects on acoustic wave measurements in observations. In Section 6, we discuss our findings.

\section{Model} \label{sec:model}

To provide a convectively stable quiet Sun background model, the method described by \citet{parch_2007_ApJ} is used. The Brunt-V\"ais\"al\"a frequency $N^2 = \frac{g}{\gamma p} \pd{p}{z} - \frac{g}{\rho} \pd{\rho}{z}$ must be positive for convective stability. Rearranging this equation in terms of $\partial \rho / \partial z$, combining with the equation for hydrostatic stability (Equation \ref{RK4_eqns2}) and introducing a free parameter $\alpha$ gives,
\begin{eqnarray}
\pd{p}{z} &=& - \rho g \label{RK4_eqns2} \\
\pd{\rho}{z} &=& -\frac{g \rho^2}{\gamma p} - \alpha \frac{\rho N^2}{g}, \label{RK4_eqns1}
\end{eqnarray}
where the gravity acceleration $g$, Brunt-V\"ais\"al\"a frequency $N$, and the adiabatic index $\gamma$ are functions of depth. The equations are solved using a fourth-order Runge-Kutta method on a one-dimensional grid. The equispaced grid covers the height range from $-50$ to $+2.48\Mm$ and is resolved with the vertical step $\Delta z=0.02626\Mm$.

To enforce convective stability, the Brunt-V\"ais\"al\"a frequency dependence on $z$ is modified by setting the negative values in the convectively unstable solar interior to small positive ones. The free parameter $\alpha$ must be greater than zero to enforce convective stability and is increased so to match the pressure in the model with the Standard Solar Model S \citep{chda_1996_Sci} pressure at a point in the interior $z = -5\Mm$. The quiet Sun model is created by smoothly joining the VAL\-IIIC \citep{vern_1981_ApJ} to Model S and integrating from the top of the solar atmosphere downwards. Using the above, a convectively stable quasi-solar model can be created with negligible change to the photospheric line formation regions and the correct sound speed profile below the surface at the expense of a slightly reduced density and pressure in the deeper regions.

Following the method described by \citet{khom_2008_apj}, an axisymmetric sunspot model is created in cylindrical geometry using three parameters $a$, $\eta$ and $B_0$ which change the sunspot radius, magnetic field inclination and strength, respectively. A full description of the effects of these parameters on the magnetic field configuration is given by \citeauthor{khom_2008_apj}. The model is defined on a two-dimensional $r$-$z$ plane discretised into a domain from $-10$ to $2\Mm$ in height, with a radius of $100\Mm$ and resolution of $\Delta z =0.1\Mm$ and $\Delta r = 0.2\Mm$.

Below $-1$ Mm depth a Low-type magnetic flux tube \citep{low_1980_soph} is constructed using an extension of the exact Schl\"uter-Temesvary formulation \citep{scl_1958_IAUS}.


For the near-surface layers $z > -1\Mm$ and in the atmosphere a Pizzo-type magnetic flux tube is used \citep{pizz_1986_apj}. The Pizzo method creates a pressure-distributed magnetic field structure through an extension of the Low formulation used above. The magnetohydrostatic equation for a non-twisted, cylindrical structure can be simplified by introducing a magnetic vector potential \citep{low_1975_apj}. This allows the problem to be reduced to a single equation for a scalar $u(r,z)$ \citep{pizz_1986_apj}
\begin{eqnarray}
\pdd{u}{r}- \frac{1}{r}\pd{u}{r}+ \pdd{u}{z} = -4 \pi r^2 \pd{p(u,z)}{u}, \label{pot_sol}
\end{eqnarray}
where $p(u,z)$ is the gas pressure along the field lines.

The Pizzo method boundary conditions require both a quiet Sun (denoted with index $q$) and umbral (denoted with index $um$) pressure, density, temperature, and pressure scale height ($h = \frac{p}{\rho g}$) and temperature distributions as functions of depth. The quiet Sun model ($p_q$,$\rho_q$,$h_q$) generated above was used for the outer boundary condition. For the inner boundary the Avrett semi-empirical model \citep{avre_1981_phss} is used, which is then joined to the pressure and density profiles at the axis of the self-similar flux tube using log-linear interpolation. This is then convectively stabilised using Equations (\ref{RK4_eqns1}) and (\ref{RK4_eqns2}) as described for the quiet Sun above. The Wilson depression can be prescribed by shifting the $\log(\tau_{5000})=0$ of the umbral model ($p_{um}$, $\rho_{um}$, $h_{um}$). The pressure and scale height are then distributed throughout the domain using the following:
\begin{eqnarray}
p(u,z) &=& p_{q}(z) - (p_{q}(z) - p_{um}(z))\left(1-\frac{u(r,z)}{u(n_r,0)}\right)^2,\label{pre_dist1} \\ 
h(u,z) &=& h_{q}(z) - (h_{q}(z) - h_{um}(z))\left(1-\frac{u(r,z)}{u(n_r,0)}\right)^2.\label{pre_dist2}
\end{eqnarray}

The potential solution given by Equation (\ref{pot_sol}) is used as an initial guess. The pressure distribution given by Equations (\ref{pre_dist1}) and (\ref{pre_dist2}) is iterated together with Equation (\ref{pot_sol}) using a Gauss-Seidel method. Thus, the complete force balance is calculated with a specified precision, giving a final distribution of the potential and pressure.

The Pizzo and Low type flux tubes are then joined at $z=-1\Mm$ and recalculated using Equations (\ref{pre_dist1}-\ref{pre_dist2}). The density and the radial and vertical components of the magnetic field vector $B_r, B_z$ are calculated according to:
\begin{eqnarray}
 \rho(r,z) = \frac{p(r,z)}{g(z) h(r,z)}\label{density} \\[4pt]
 B_r(r,z)= -\frac{1}{r}\pd{u}{z} \label{mag_field_r} \\
 B_z(r,z) = \frac{1}{r}\pd{u}{r}. \label{mag_field_z}
\end{eqnarray}
To extend this model below $z = -10\Mm$ a vertical flux tube with a constant $B_z$ and zero $B_r$ is used, and the pressure and density profiles are continued smoothly downwards.

Finally, the FreeEOS equation of state \citep{freeeos} is applied to find the adiabatic index, temperature and sound speed at each grid cell in the model. The model is then converted to Cartesian geometry, giving the full set of physical parameters required for the MHD simulations and radiative transfer calculations.

\begin{figure}
\center
\plotone{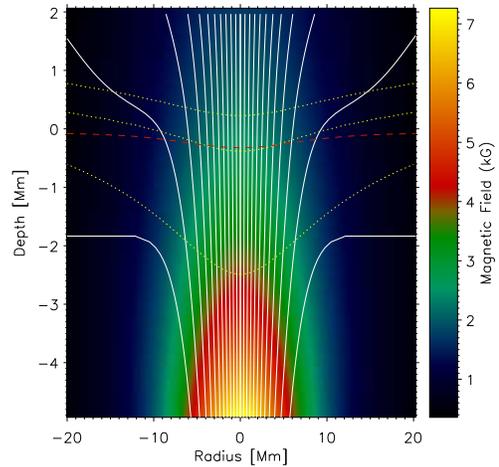}
\caption{Magnetic field structure of the sunspot model. The magnetic field strength is shown with the magnetic field lines overplotted (solid). Also shown are the ${c_s}/{v_A} = 1$ (middle dotted), 0.1 (upper dotted) and 10 (lower dotted) contours. The dashed line is the $\log(\tau_{5000})=0$ contour, representing the visible photosphere. Note that the aspect ratio is severely stretched.}
\label{fig1}
\end{figure}

Using the procedure explained above the magnetic field structure pictured in Figure \ref{fig1} was constructed. The background image in the figure shows the modulus of magnetic field $B$. The field lines are nearly vertical in the ``umbral" region ($r<10~\mathrm{Mm}$), and show inclination of about $60^{\circ}$ in the ``penumbral" region, $r>10~\mathrm{Mm}$, of the sunspot model. 

In the figure, the dashed line shows the $\log(\tau_{5000})=0$ layer, while the dotted contours represent $c_s/v_A=0.1,~1,~\mathrm{and}~10$ levels. As is evident from the figure, in the umbral region at the axis of the sunspot, the $\log(\tau_{5000})=0$ layer is positioned higher than $c_s/v_A=1$ layer, suggesting formation of the continuum radiation in the magnetically-dominated sunspot atmosphere.

\begin{figure}
\epsscale{0.7}
\center
\plotone{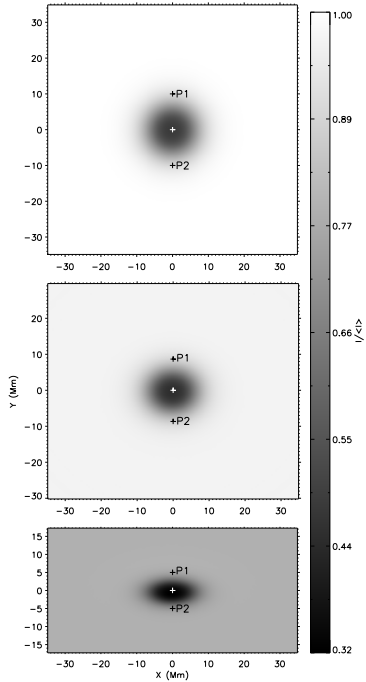}
\caption{$6173\mathrm{\AA}$ continuum intensity of the sunspot model at the observational angles (top to bottom) $0^{\circ}$, $30^{\circ}$ and $60^{\circ}$ to the vertical. The figures have been normalised to the quiet Sun value at $0^{\circ}$ inclination. Inclination is performed towards an observer displaced in the negative $y$ direction. The crosses show the two penumbral, and one umbral point used in Figure \ref{fig3}}
\label{fig2}
\end{figure}

The 6173\AA~photospheric absorption line of neutral iron is used for observations of the full solar disk with the Helioseismic Magnetic Imager (HMI) onboard the Solar Dynamic Observatory (SDO). Therefore, this line was chosen to carry out radiative diagnostics of the sunspot model using the SPINOR code \citep{sol_1987_phd, shel_2007_AA}. For each one-dimensional column of the model, continuum intensity and spectral line profile calculations are performed by solving the Unno-Rachovsky \citep{unno_1956_pasj} radiative transfer equation for the Stokes vector $I = [I,V,Q,U]$. Off-disk centre observations are simulated by inclining the numerical domain and interpolating the density, temperature, magnetic field and velocities onto the new line of sight ({\it los}). The slanting is performed around the $z = 0 \km$ height and in the direction of positive $y$ (Figure \ref{fig2}). The velocity and magnetic field vectors are then projected into the new reference frame. The calculation uses 500 wavelength points with a $\delta \lambda = 0.002\mathrm{\AA}$ to ensure the spectral line is highly resolved. The {\it los} velocity is given by $v_{los}=v_z \cos\theta + v_x \sin\theta$. The magnetic field is recalculated using a similar relation.

Figure \ref{fig2} shows the continuum images of the sunspot model calculated for $0^{\circ}$, $30^{\circ}$ and $60^{\circ}$ angles between the surface and the {\it los}, which correspond to viewing cosine $\mu=1.,~0.866~\mathrm{and}~0.5$, respectively. We find that the model produces a realistic limb darkening dependence with a continuum value of 79\% of the disk centre intensity at $\mu = 0.5$. This is only slightly higher than the 75\% of the limb darkening curve determined by \cite{fou_2004_ApJ}.

\epsscale{1.}
\begin{figure*}
\center
\plotone{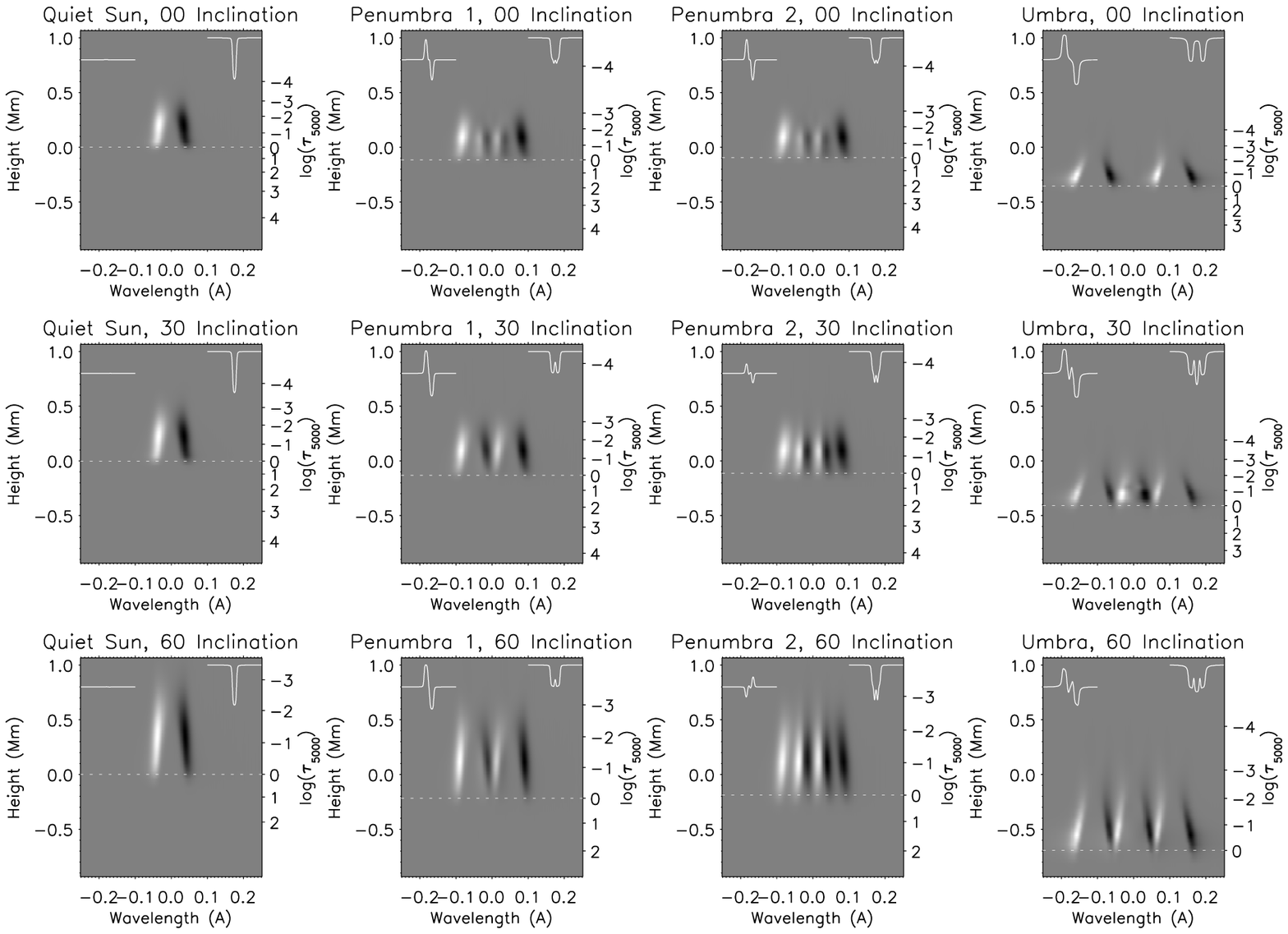}
\caption{{\it los} velocity response functions of Stokes $I$ profile of $6173\mathrm{\AA}$ Fe{\sc i} line computed for the models of quiet Sun (first column), near-side and far-side penumbrae (second and third columns, respectively), and umbra (fourth column) for $\mu=1.,~0.866~\mathrm{and}~0.5$ positions at the solar disk.
The y-axis represents the height along the {\it los} at which the perturbation is placed, where $0\Mm$ represents the $\log\tau_{5000}=0$ layer for the quiet Sun photosphere. The $5000\mathrm{\AA}$ optical depth axis is also shown, with a dashed line showing the observed depression of the photosphere. The corresponding Stokes-$I$ (top right) and Stokes-$V$ (top-left) profile shapes were over-plotted in white in each panel over the wavelength range shown in the figure.
}
\label{fig3}
\end{figure*}

The velocity response functions of the $6173\mathrm{\AA}$ spectral line are shown in  Figure \ref{fig3} for the quiet Sun, two penumbral regions at $\pm 10 \Mm$, and in the centre of the sunspot umbra for the chosen positions at the solar disk. These locations have been marked with crosses in Figure \ref{fig2}. Since, for observations away from solar disk centre, two points at the same distance from the sunspot axis are not equivalent, the penumbral models have been chosen so that {\it los} of P1 crosses the umbral region, while the {\it los} of P2 inclines further into the penumbra. The response functions were calculated by computing a perturbed profile with a small positive (directed towards the observer) {\it los} velocity perturbation and subtracting from it an unperturbed profile for the same location.

The top row of Figure \ref{fig3} shows the response functions of the four points in the model for the disk centre calculation. The perturbation is directed towards the observer, causing the line to be blue-shifted. The lobes of the response function tilt inwards towards the line core, marked as $0.0\mathrm{\AA}$ in the plots, clearly demonstrating dependence of the sensitivity of the profile on height; the regions closer to the line-core (on the $x$-axis) are formed higher in the atmosphere. 

The top-right figure shows a fully Zeeman-split profile in a strong umbral magnetic field. Notably, while the line formation height range is narrower compared to the quiet Sun, the response function lobes are wider in wavelength, suggesting higher sensitivity of the line to velocity perturbations.

The two penumbral points are identical in the solar disk centre simulation due to the symmetry of the sunspot. As the penumbral magnetic field is weaker, the line is not completely split. In the line core, the response function shows two smaller regions of sensitivity to the velocity perturbation.

Figure \ref{fig3} demonstrates that the line formation height range increases with the inclination angle. In the case of the quiet Sun (left column of the figure), it increases from $\sim400\km$ at the disk centre to $\sim800\km$ at $\mu=0.5$. As the line width does not change significantly, the wavelength range of the response function does not change with the inclination.

For the cases of magnetised penumbral and umbral atmospheres, the observed visible sunspot surface increases with the inclination angle. Between $\mu=1$ and 0.5, the $\log\tau_{5000}=0$ level for the penumbral points P1 and P2 is shifted downwards by $\sim100\km$, and by $\sim400\km$ for the umbra. The line sensitivity height range also increases further away from the disk centre, similarly to the quiet Sun.  

Notably, the line profiles and the response function shapes for P1 and P2 are very different. The far-side umbra (second column of Figure \ref{fig3}, P1 in Figure \ref{fig2}) will have a formation range that extends into the highly magnetic umbra. This can be observed as an increasingly split profile as inclination increases in the 2nd and 3rd rows. The near-side penumbral pixel (third column, P2) will similarly form in a region of lower magnetic field. Due to the inclination of the magnetic field, P1 will measure a higher magnetic field strength along the {\it los} while P2 will incline against the direction of field line inclination.
 
The angle of the two ridges is seen to be larger in the umbral distributions than the quiet sun. For a small wavelength range in the quiet sun up to $500 \km$ of the atmosphere will be measured. Comparitively, in the umbral distribution a similar filter would only sample around $100 \km$.

The large range of responses seen in the different penumbral and umbral positions will lead to larger uncertainty in the observation height of velocity measurements. The impact of Zeeman-split profiles on velocity measurements is only amplified at higher inclinations.

\section{Numerical Simulations}

We perform a simulation of acoustic wave propagation in the simulated model with the SPARC code. The code was designed to solve the linearised ideal MHD equations for wave propagation in a stratified solar environment \citep{hana_2011_ascl}. The version of the code we employ for the simulations uses Message Passing Interface (MPI) to parallelise the computation and reduce computation time. It uses an implicit compact 6th order finite difference scheme applied to the horizontal and vertical derivatives.  An explicit filter is used to prevent numerical instabilities in the solution.
 The boundary conditions used in the simulation include a Perfectly Matched Layer (PML) \citep{hana_2010_aa} at the top and bottom boundaries, allowing for efficient absorption of the outgoing waves. A $12.5~\mathrm{Mm}$ `sponge' type absorbing layer is used on the side boundaries, which adds a linear friction term to the governing equations \citep{colo_2004_AnRFM}. 
The code includes a Lorentz force limiter \citep{remp_2009_ApJ}, which is required due to the high Alfv\'en speed in the solar atmosphere. The limiter, although unphysical, prevents reduction of the time step and excessive computational times. The Alfv\'en speed limiter is set at $v_A = 125~\mathrm{km\,s^{-1}}$, which is sufficiently high to allow fast MHD waves of interest, which have horizontal phase speed less than this, to propagate and refract correctly while still allowing us to use a manageable time step. Our cap is large enough for this to not be an onerous constraint. The implications of the limiter on helioseismic travel time shifts have been studied in detail by \cite{mora_2014_ApJ}.

The numerical grid has horizontal extent of $n_x = n_y = 256$ grid points, with a physical size of $140\Mm$, giving a resolution of $\Delta x = \Delta y = 0.55~\mathrm{Mm}$ in the horizontal directions. To deal with the large variation in physical parameters over the domain from the convective zone to chromosphere, the code uses a vertical grid spacing based on the sound speed. The grid has $n_z = 300$ points between $1.5$ and $-25~\mathrm{Mm}$ and is distributed such that the acoustic travel times between each cell are the same for the quiet Sun, $\Delta z \propto {1}/{c_s}$. This gives a resolution of around $50~\mathrm{km}$ near the photosphere, and around $1~\mathrm{Mm}$ in the lower solar interior. This means we do not resolve slow waves in the large $\beta$ regime, but these are effectively decoupled from the system anyway, so their neglect is not important. The following acoustic source, similar to that described by \citet{shel_2009_AA}, was used: 
\begin{eqnarray}
v_z = A_0 \sin(\frac{2 \pi t}{T_o})\exp{\frac{-(t-T_1)^2}{\sigma^2_t}}\exp{\frac{-(r-r_0)^2}{\sigma^2_r}}\times \label{pulse}\\
\times\exp{\frac{-(z-z_0)^2}{\sigma^2_z}}, \nonumber
\end{eqnarray}
where $T_0 = 300$ s, $T_1 = 600$s, $\sigma_t = 100$ s, $\sigma_{xy} = 1$ Mm, $\sigma_z = 0.25\Mm$. The position of the pulse is $r_0(x,y)=(45,70)\Mm$, $z_0 = -0.65\Mm$.

The SPARC code solves the MHD equations for the perturbations around the MHS background model. A master-slave Open-MPI code has been written to take these perturbations, combine them with the background model and incline them as required. The SPINOR routines are then applied to each pixel to generate the full Stokes vector for each pixel. Using the generated Stokes-$I$ profiles, the {\it los} centre-of-gravity Doppler velocity is calculated by computing the position of the centre of gravity of the line profile and determining its shift from the unperturbed counterpart, computed for the background model, according to:

\begin{equation}
\Delta \lambda = \lambda_{cog} - \lambda_0 = \frac{\int (I_c-I) \lambda d\lambda}{\int (I_c-I) d\lambda}-\lambda_0.
\label{linecog}
\end{equation}

To calculate bisector Doppler velocities from the spectral line the relative intensity $I_{rel}$ was determined by normalising the measured Stokes I between 0 and 1. Bisectors of the spectral line were calculated at 100 evenly spaced values between $0.05-0.95$ of $I_{rel}$. The bisectors were calculated for the background model and for each output snapshot. A Doppler velocity was then determined for each snapshot using the shift from the unperturbed background value, according to:

\begin{equation}
v_{bsr} = (\lambda_0 - \lambda_{bsr}) \frac{c}{\lambda_0}. 
\label{bsrvelocity}
\end{equation}

\goodbreak\section{Results}

\begin{figure}
\center
\plotone{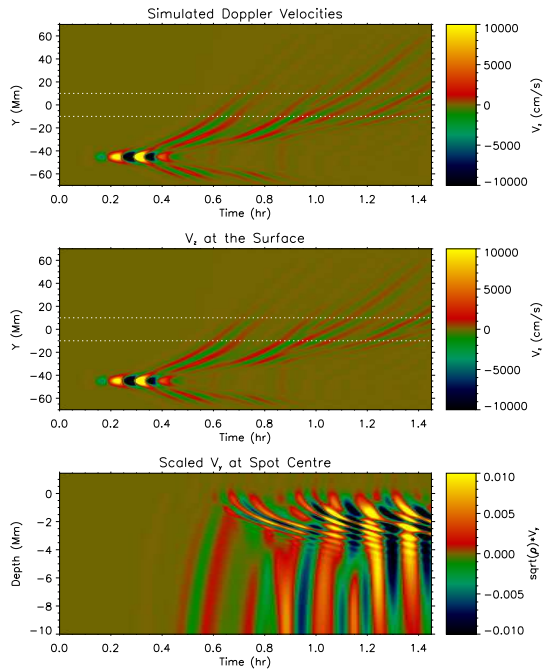}
\caption{Response of the model to the acoustic source. Top panel - simulated Doppler velocities at $0^{\circ}$ inclination, measured at $x=0$. Middle panel - simulated vertical component of velocity at a geometric height $z=0\Mm$, measured at $x=0$. The differences between these two can be seen to be small. Bottom pannel - $\sqrt\rho v_y$ through spot centre (0,0) Mm showing the propagation of slow modes down through the box. Two dashed horizontal lines in the top plots mark the position of the sunspot umbra.}
\label{fig4}
\end{figure}

Using the model described in Section 2 and methodology given in Section 3, a $2.5$ hour simulated observation of wave propagation through the sunspot model was performed. The top panel in Figure \ref{fig4} shows a time-distance plot of the centre-of-gravity {\it los} Doppler velocity measured using Equation (\ref{linecog}). The first three wave bounces can be easily seen. A shift in the wave arrival time can be observed as a flattening of the wavefront as it passes through the sunspot umbra at $y=0\Mm$. The middle panel of Figure \ref{fig4} shows a time-distance plot of the vertical component of velocity at the $z=0\Mm$ level of the simulation domain.  A comparison between the top two panels shows that the vertical velocity in the domain and the {\it los} Doppler velocity are visually identical. Some reflection can be seen from the top and bottom PMLs, and from the side boundaries.

The bottom panel of Figure \ref{fig4} shows the horizontal component of velocity, scaled by $\sqrt{\rho}$ to provide a view of the slow magnetoacoustic wave in the strong magnetic field. A slice is taken through the centre of the simulated sunspot $(x=0, y=0)$. The fast wave can be seen to propagate through the sunspot in the lower interior where plasma-$\beta$ is high. At around $z=-0.400\Mm$ in the umbra, the incoming fast wave hits  ${c_s}/{v_A} = 1$ layer (See Figure \ref{fig1}) and undergoes partial transmission as a slow mode (effectively acoustic in $c_s<v_A$). The slow magnetoacoustic wave (now magnetic in $c_s>v_A$) can be seen to propagate back down into the sunspot as a flattening banding in the time distance plot.  The wave amplitude in the atmosphere is low due to scaling by the very low densities, however, it still can be seen to continue to travel upwards above the photosphere and escapes through the absorbing upper boundary.

\begin{figure}
\center
\plotone{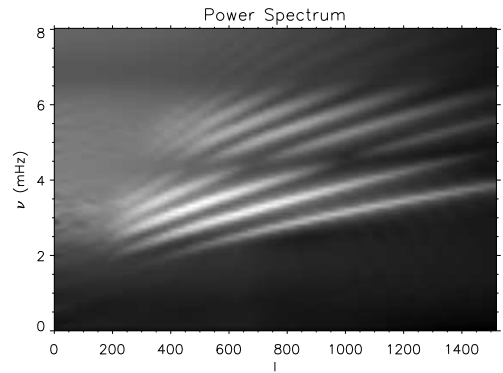}
\caption{A $\nu-l$ power spectrum for the sunspot box calculated from synthetic Doppler velocities.}
\label{fig5}
\end{figure}

Figure \ref{fig5} shows a power spectrum plotted with azimuthally averaged wavenumber and frequency. The spectrum has been calculated from the $2.5$ hour time-series of centre-of-gravity velocities calculated from the synthesised line profile. The power ridges are well resolved, although there are gaps in power at $4.5~\mathrm{mHz}$ and shifts seen for high $l$. Similar gaps are found in the simulations of \citet{parch_2007_ApJ} with high top boundary (1.75 Mm; their Fig. 6c), and attributed to trapping of acoustic modes. We do not understand how acoustic trapping explains this phenomenon. The gaps are also present in quiet sun simulations (no magnetic field), but are largely removed when the top boundary is lowered to 500 km above the solar surface (their Fig. 6b). This suggests that there is some numerical dissipation mechanism operating in our model chromosphere ($z > 500 \km$) that we are yet to identify.

Using the Fe{\sc i} $6173~\mathrm{\AA}$ spectral line data, velocity shifts were calculated for each bisector as the differences between the perturbed and background values (Equation (\ref{bsrvelocity})). From the 2.5 hour time series of these velocities acoustic power maps of wave propagation in the sunspot were created. The power maps are calculated for 100 bisector depth positions by taking the fast Fourier transform of the velocity time series for each of the $(x,y)$ pixels in the model. The acoustic power is binned into different frequency bins by applying a Gaussian filter with a FWHM of $0.5~\mathrm{mHz}$ around the chosen central frequency. For each frequency band and inclination the acoustic power is normalised by its average in the simulation domain. 

By taking the Doppler velocity measurements at multiple bisector depths, it was possible to disentangle the dependence of the wave behaviour on the height within the line formation region. As Figure \ref{fig3} shows, bisectors taken higher in the line profile are formed deeper and closer to the continuum formation height, at a physical height of around $150\km$, while bisectors taken deeper are closer to the absorption line core and formed higher at a physical height of around $500\km$. As was already noted, these formation heights vary significantly in the penumbra and umbra of the sunspot, with an offset of $350\km$ due to the prescribed Wilson depression.

\begin{figure*}
\center
\plotone{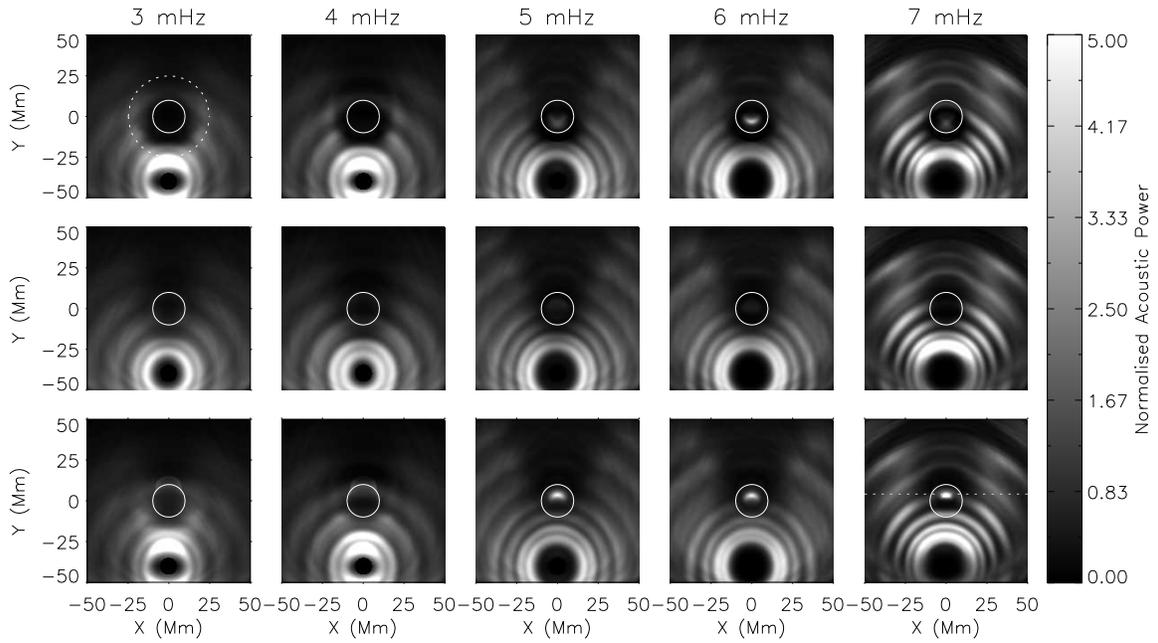}
\caption{Acoustic power map calculated from the shifts in the bisector of the Fe{\sc i} 6173\AA~line at a bisector height of $0.3 ~I_{rel}$. The columns, from left to right, show power in the 3, 4, 5, 6 and 7 mHz bands. The rows, from top to bottom, show measurements made at $-60^{\circ}$, $0^{\circ}$ and $60^{\circ}$ inclination from the vertical, where the field of view has been inclined in the $y$-direction. The power at each inclination angle has been normalised by its average in each frequency band. This represents a velocity measurement made near the line-core, showing the behaviour higher in the atmosphere. The sunspot umbra has been marked with a solid circle, while the dashed line in the bottom right panel represents the slice taken in Figure \ref{fig8}. The low frequency ring has been marked in the top left panel.}
\label{fig6}
\end{figure*}

\begin{figure*}
\center
\plotone{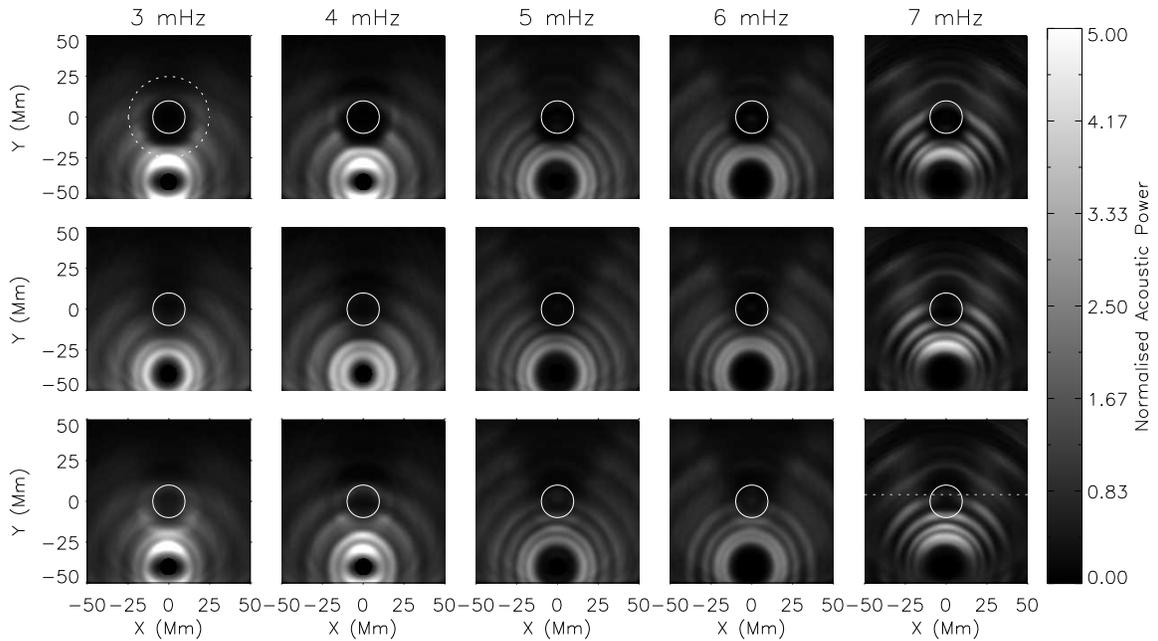}
\caption{Acoustic power map calculated from the shifts in the bisector of the Fe{\sc i} 6173\AA~line at a bisector height of 0.9 ~$I_{rel}$. The layout of the columns and rows is as seen in Figure \ref{fig6}. This figure represents measurements made near the line wings.}
\label{fig7}
\end{figure*}

First we aim to investigate the horizontal distribution of acoustic power throughout the simulation domain. The region of the acoustic source at $x=45\Mm$, $y=70\Mm$ has been masked in the power maps. Figures~\ref{fig6} and~\ref{fig7} show the acoustic power calculated from the Doppler shifts measured at the bisector positions of $0.3~I_{rel}$ and $0.9~I_{rel}$, respectively. The acoustic power was binned into frequency 1 mHz bands centred at 3, 4, 5, 6 and 7 mHz (left to right in the figures), and the disk positions of $-60^{\circ}$, $0^{\circ}$ and $60^{\circ}$ were used (top to bottom). {\it los} velocity measurements off the disk centre are affected by a larger line formation region and the presence of horizontal velocity components in the {\it los} velocities. Immediately obvious in the figure is a series of concentric rings of power travelling out from the source. These rings can be thought of as the spatial analogue of the ridges seen in Figure \ref{fig5} and occur at discrete values because a single point-like source was used in the simulation.

The differences between the acoustic power maps in Figure \ref{fig6} and~\ref{fig7} represent changes in the wave behaviour over the formation height of the spectral line. This formation height can span almost a megameter close to the solar limb (see Figure \ref{fig3}), and the differences are substantially more pronounced in the magnetic regions. 

In the acoustic power maps (Figure \ref{fig6} and \ref{fig7}) a solid line representing the sunspot umbra is over-plotted. Outside the sunspot umbra the sunspot shadow is observed at all frequencies. In the sunspot shadow, the power from the pulse has been absorbed or reflected by the magnetic field structure. This is most obvious at lower frequencies (left two columns). As frequency increases, the ridges of increased acoustic power can still be seen behind the sunspot. Interestingly, the magnetic field perturbs the concentric rings seen at 7 mHz, as the fast wave propagation speed and turning height change. Behind the sunspot in the $7~\mathrm{mHz}$ band (right column), a small region of increased power can be seen between the two outermost rings of the power ridges (around $y = 30\Mm$). As this is seen in high frequencies and as a ring around the sunspot, rather than around the source, it appears to be a far-side acoustic halo. Acoustic halos are seen around active regions as an excess of acoustic power compared to the quiet sun. \footnote{A more comprehensive look at physics of acoustic halos, based on three-dimensional simulations of this sunspot model can be found in \citet{rijs2015}, which expands on the previous works of \citet{hana_2008_ApJ,khom_2009_AA}. They are attributed to fast MHD waves reflected in active region atmospheres by the steep Alfvén speed gradients there.}
 
Comparison of the weakly-magnetic regions of each column in Figure \ref{fig6} shows little variation with inclination, regardless of frequency. In these regions the propagating fast wave dominates the simulated observations as there is little magnetic field to alter the fast-wave turning point or to allow for mode-conversion process to take place. Inside the umbral region, the acoustic power maps for the disk centre case show very little power in 3 and 4 mHz, and the power in the vertically-aligned oscillations is seen to be almost completely absorbed. As the inclination increases, the 3 and 4 mHz power in the umbra remains low, while the 5, 6 and 7 mHz power bands show a significant power enhancement.

From the response function (first column in Figure~\ref{fig3}) we expect there to be larger differences in power between the line core (Fig.~\ref{fig6}) and line wings (Fig.~\ref{fig7}) as we observe further away from the solar disk centre. There is little difference found in the disk centre cases (top row) between the two figures. However, at $60^{\circ}$ inclination significant differences can be seen between the power maps at the line core and line wings. Particularly, (1) the umbral power increase is only seen at the line core (Fig.~\ref{fig6}); (2) the ring-like structure (marked by the dashed circle in the left panel of Fig.~\ref{fig6}) found at around $y=-20\Mm$ in the bottom left two panels is somewhat wider at the line core than in the line wings (Fig.~\ref{fig7}). This structure is most apparent in the $3-4~\mathrm{mHz}$ frequency bands (first two columns), and the power in it decreases with increasing frequency. While the inclination of the magnetic field at the surface at the radius of $20\Mm$ is $60^{\circ}$ degrees from vertical the magnetic field strength is low, and the equipartition layer $c_s=v_A$ is located above the line formation region. Therefore, the ring is of acoustic nature and cannot be related to the slow magneto-acoustic mode, as it is found at the source side in both $-60^{\circ}$ and $60^{\circ}$ inclinations corresponding to the {\it los} direction which is either parallel or highly inclined to the magnetic field.


As demonstrated, the umbral and penumbral acoustic power structures are mostly seen near the line core (Figure \ref{fig6}). In the {\it los} velocities measured from bisectors in the line wings (Figure \ref{fig7}) only a faint structure can be seen at high inclinations, again more obvious in the 7 mHz power band (bottom right).

\begin{figure*}
\center
\plotone{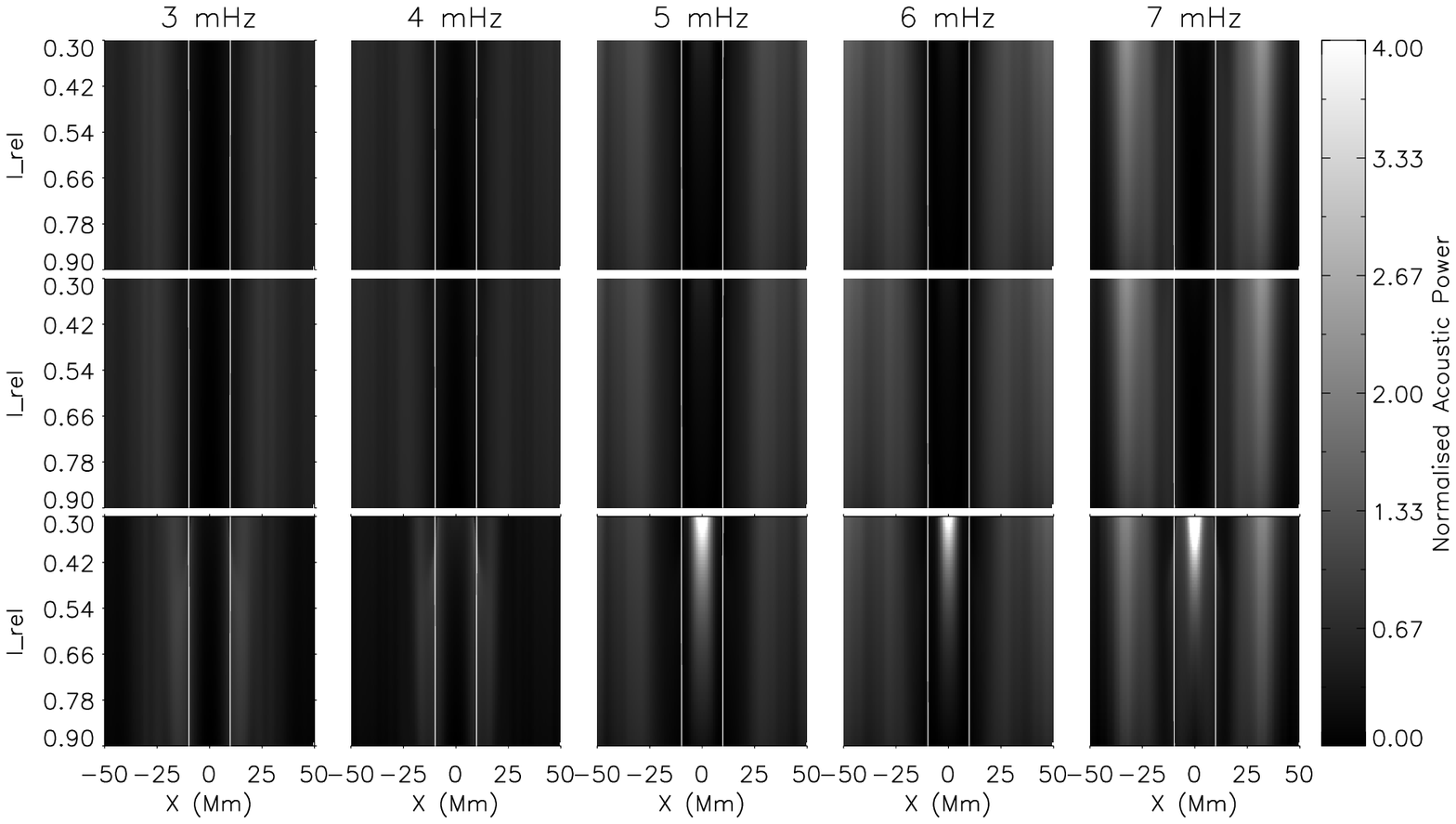}
\caption{Bisector power map at x = 0 Mm. The layout of the columns is as in Figure \ref{fig6}. The rows, from top to bottom, show measurements made at $0^{\circ}$, $30^{\circ}$ and $60^{\circ}$ inclination from the vertical. The Y axis in this figure represents the line depth where the bisector wavelength and Doppler velocity are measured and covers a large part of the line formation region of $400-800 \km$ in length. The formation region will depend on the magnetic field strength and inclination (Figure \ref{fig3}).}
\label{fig8}
\end{figure*}

To better understand the three-dimensional structure of the umbral power increase, in accordance with the response functions shown in Figure \ref{fig3}, a multi-height observation is made by computing the Doppler velocities from bisector shifts measured at different line depths within the line formation region. In Figure \ref{fig8}, a Doppler velocity map for a slice marked by a dashed line in Figure \ref{fig7} is plotted for each bisector depth in the range $0.3-0.9 ~I_{rel}$. 

For the $0^{\circ}$ and $30^{\circ}$ inclinations (top and middle rows of Figure \ref{fig8}), the changes in the structure and magnitude of acoustic power over the line formation range are limited to an increase in power in the high frequency bands for observations made closer to the line core. This matches the observations of acoustic halos by \citet{raj_2013_SoPh}, where the acoustic power was weaker using filtergrams close to the line-wings. At $60^{\circ}$ inclination (bottom row, left two columns) the faint $x=\pm 20~\Mm$ radius low frequency ring can be made out, increasing in radius at larger heights.

The vertical extent of the umbral power structure, seen at high inclinations at 6--7 mHz (bottom right two panels in Figure \ref{fig6}--\ref{fig8}) shows a significant increase in power higher in the formation region. The formation of this acoustic power structure seems to start mid-way up the line formation region, suggesting a highly localised phenomenon. The inclination of the field lines at the centre of this power increase ($y = 4.11~\Mm$) is approximately $20^{\circ}$ from the vertical. Taking into account the observation angle of $\pm 60^{\circ}$, the field line is almost perpendicular to the line-of-sight. Comparison with Fig.~\ref{fig1} shows the continuum formation height of the spectral line and the ${c_s}/{v_A}=1$ layer cross at $z = 0~\Mm$ and $4~\Mm$ radius from the sunspot centre, allowing for direct observation of conversion of fast (parallel to the magnetic field line, perpendicular to the {\it los}) waves to slow (perpendicular to the magnetic field line, and parallel to the {\it los}) waves. The increased power in this region corresponds to the slow magneto-acoustic wave in the region where the magnetic field is close to perpendicular to the {\it los}. This result confirms previous findings by \citet{zhar_2013_AnGeo} that the observed acoustic power increase in the sunspots is a signature of slow magneto-acoustic waves.

\section{Discussion and Conclusion}

In this paper we have: (1) described a modification of the Khomenko sunspot model we developed to provide a more accurate line formation region, allowing for accurate spectral synthesis to be performed; (2) analysed response functions in our model to understand our synthesised centre-to-limb observations of the Fe{\sc i} $6173\mathrm{\AA}$ spectral line in a model sunspot; (3) have investigated wave propagation through the model using a three-dimensional simulation of magnetoacoustic wave propagation; (4) computed the spectral line profiles and provided a time series of the simulated observations of the sunspot model using the simulation data; (5) used spectral line bisector measurements to perform multi-height simulated observations over the line formation region; (6) studied maps of the acoustic power in the sunspot to better understand the absorption line response to the oscillations in sunspots and the effects of non-locality of radiative transport on helioseismic measurements.

The sunspot is found to absorb or scatter the incoming acoustic wave energy in all but the 6 and 7 mHz frequency band. There are signatures of a slight power enhancement seen around the sunspot, similar to those seen in observations of acostic halos.

Small ridges of acoustic power can be seen in the 7 mHz band in the shadow of the sunspot. \citet{zhar_2013_AnGeo} showed in a simple 2D simulation that there was a significant power enhancement as a far-side acoustic halo. We can see slight power enhancements in the range of 25-40 Mm from the sunspot. Comparing the different rows in Figures \ref{fig6} and~\ref{fig7} shows that outside the sunspot umbra, the horizontal and vertical velocities behave similarly, with only minor differences in the power maps.

The appearance of a high-frequency anomalous acoustic power excess in the sunspot centre -- the umbral ``belly button" -- can be seen predominantly in the case of $60^{\circ}$ inclination with faint traces at lower inclinations. This geometry suggests that it is driven by slow magneto-acoustic waves, as it is seen when the horizontal velocity component dominates the {\it los} velocity (bottom rows Figure \ref{fig6} and~\ref{fig7}). It can be seen as a crescent-like structure, which is most dominant towards the line core (higher in the atmosphere, Figure \ref{fig6}) and very faint in the spectral line wings (lower in the atmosphere, Figure \ref{fig7}). Umbral power enhancements are seen in the space-based (HMI, \citet{zhar_2013_AnGeo}) and ground based \citep{balt_1998_SoPh} observations of sunspots. Notably, no power excess was observed in the G-band \citep{naga_2007_PASJ}. The appearance of this power increase in a simulated sunspot suggests a magneto-acoustic phenomenon, rather than photon noise \citep{donea2015}. There are many differences in both sunspot properties and the radiative effects on HMI measurements that could explain the lack of the umbral power increase in {\it all} acoustic observations of sunspots. These include changes in the Wilson depression, the wide range of the velocity-response function in a magnetic structure (Figure \ref{fig3}), low resolution and high-noise measurements off the disk-centre or issues with using discrete filters on highly split profiles.

Current measurements of acoustic travel times in computational helioseismology are largely performed using measurements at the geometric heights in the simulation domain \citep{mora_2013_JPhCS}, or on a surface roughly representing the continuum formation height determined by the $\log(\tau_{5000})=1$ layer in the simulation domain \citep{khom_2012_ApJ,mora_2015_mnres}. Despite the fact that the physical velocity in the simulation matches reasonably well to the {\it los} velocities calculated from the simulated spectral lines, this method misses a lot of information that can otherwise be gained from the range of formation of the spectral lines. As we show, this range also changes substantially if the simulation is performed for positions at the solar disk away from the centre. Using the model we described, artificial observables mimicking the HMI and MDI pipelines can be made \citep{sche_2012_SoPh,flec_2011_SoPh}, as well as comparisons to ground based observations.

The multi-height Doppler measurements made by \cite{naga_2014_SoPh}, using the HMI filter-grams provide a similar approach to multi-height measurements as the bisectors used in this study. Rather than velocity measurements made using shifts in Stokes-$I$, HMI uses measurements of both Stokes $I+V$ and Stokes $I-V$. As the next step, it will be important to fully simulate the HMI data pipeline response to a variable magnetic atmosphere before a direct comparison to the HMI measurements can be made.

\acknowledgements

Sergiy Shelyag's research is supported by Australian Research Council Future Fellowship. Damien Przybylski and Sergiy Shelyag's research is performed with the computational resources provided by the Multi-modal Australian ScienceS Imaging and Visualisation Environment (MASSIVE) (www.massive.org.au) and through Astronomy Australia Limited by the NCI National Facility systems at the Australian National University and the Centre for Astrophysics and Supercomputing of Swinburne University of Technology (Australia).

\end{document}